\documentclass[10pt, conference, compsocconf]{IEEEtran}

\usepackage{cite}
\usepackage[cmex10]{amsmath}

\usepackage{multirow}
\usepackage{amsmath}
\usepackage{amssymb}
\usepackage{graphicx}
\usepackage[tight,footnotesize]{subfigure}

\usepackage{color}
\usepackage{url}
\usepackage{array}

\newcommand{\Reals}{{\mathbb R}}             

\newcommand{\loss}{L}

\newcommand{\HilbertSpace}{\mathcal{H}}
\newcommand{\tsize}{n}
\newcommand{\Tset}{T}

\newcommand{\kernel}{k}
\renewcommand{\vec}[1]{\mathbf{#1}}
\newcommand{\ip}[2]{\langle #1 , #2 \rangle}
\newcommand{\norm}[1]{{\left|\left|#1\right|\right|}}

\newcommand{\minimize}{\operatornamewithlimits{minimize}}

\newcommand{\ttt}[1]{\texttt{#1}}

\begin{document}

\title{Detecting Quasars in Large-Scale Astronomical Surveys}

\author{\IEEEauthorblockN{Fabian Gieseke\IEEEauthorrefmark{1}, Kai Lars Polsterer\IEEEauthorrefmark{2}, Andreas Thom\IEEEauthorrefmark{1}, Peter Zinn\IEEEauthorrefmark{2},\\Dominik Bomanns\IEEEauthorrefmark{2}, Ralf-J\"urgen Dettmar\IEEEauthorrefmark{2}, Oliver Kramer\IEEEauthorrefmark{3}, and Jan Vahrenhold\IEEEauthorrefmark{1}}

\IEEEauthorblockA{\IEEEauthorrefmark{1}
Faculty of Computer Science, Technische Universit\"{a}t Dortmund, Dortmund, Germany\\
\{Fabian.Gieseke, Andreas.Thom, Jan.Vahrenhold\}@tu-dortmund.de}
\IEEEauthorblockA{\IEEEauthorrefmark{2}
Department of Physics and Astronomy, Ruhr-University of Bochum, Bochum, Germany\\
\{Polsterer, Zinn, Bomans, Dettmar\}@astro.rub.de}
\IEEEauthorblockA{\IEEEauthorrefmark{3}
International Computer Science Institute, Berkeley, USA\\
okramer@icsi.berkeley.edu}
}


\maketitle

\begin{abstract}
We present a classification-based approach to identify quasi-stellar radio sources (quasars) in the Sloan Digital Sky Survey and evaluate its performance on a manually labeled training set. While reasonable results can already be obtained via approaches working only on photometric data, our experiments indicate that simple but problem-specific features extracted from spectroscopic data can significantly improve the classification performance. Since our approach works orthogonal to existing classification schemes used for building the spectroscopic catalogs, our classification results are well suited for a mutual assessment of the approaches' accuracies.

\end{abstract}


\section{Introduction}
The automated analysis of data sets has become an increasingly important issue for researchers in astronomy~\cite{BallB2009,Borne2009}. This is in particular the case for massive data sets obtained from, e.g., the \emph{Sloan Digital Sky Survey} (SDSS), which is said to be ``one of the most ambitious and influential surveys in the history of astronomy''~\cite{sdss2010}. The latter catalog is currently based on raw data of about 60~terabytes and the trend towards data-intensive science seems to become even more evident with near-future projects such as the \emph{Large Synoptic Sky Telescope}~\cite{lsst2010}, which will produce data volumes in the petabyte range. From a machine learning perspective, a variety of problems in astronomy can be formulated as supervised (e.g. classification, regression) or unsupervised tasks (e.g. clustering, dimensionality reduction), and the corresponding tools for addressing these tasks have been recognized as ``increasingly essential in the era of data-intensive astronomy''~\cite{Borne2009}.

We describe the use of supervised learning techniques to discriminate quasars from other celestial objects. We approach this problem from a machine learning point of view and analyze how the generalization performance of standard classifiers can be improved by incorporating simple problem-specific features which are motivated by the physical properties of quasars and other celestial objects. Although some classification problems have already been addressed in the field of astronomy, the interface between both communities is not well defined~\cite{Borne2009}. One of the goals of this paper is to work towards bridging the gap between both communities by describing an important astronomical classification problem (along with some physical background) and by preparing the corresponding data such it can be approached easily from a machine learning point of view.
\section{Background}
\label{sec:background}
One of the main objectives of the SDSS consisted in finding \emph{quasi-stellar radio sources (quasars)}, a special kind of \emph{Active Galactic Nuclei} (AGN). Due to their extreme luminosities, quasars are among the most distant objects in the universe that can be observed. Depending on this distance it takes up to billions of years for the emitted radiation to reach the Earth. Therefore this radiation reveals information about the long ago state of a quasar and thus about the early universe.

The widely accepted explanation for the nature of these extreme sources that reach the highest luminosities among all known
astrophysical phenomena is the so-called \emph{unified model}~\cite{Urry1995}. This model reduces the large number of different types of AGN to a single phenomenon: a \emph{Supermassive Black Hole} (SMBH) surrounded by a thick dust torus that is accreting material. It is heated to extremely high temperatures \cite{Bonning2007} and thus produces radiation in nearly all frequency bands. The differences in how we observe AGNs is mainly caused by the inclination of the surrounding dust torus. If the inclination angle is small, we are able to observe the inner parts of the torus. In this case either an optically bright quasar or even a highly variable blazar is detected. The broad emission lines in the spectrum of a quasar (see below) result from the fact that, while observing the direct vicinity of the SMBH, one observes a region with a higher gravitational potential and thus higher orbiting velocities of the accretion disk material. Such high velocities yield to a Doppler-broadening of the emission lines, in this strength only present for AGN-powered sources.

\paragraph{Related Work}
The problem of identifying quasars has already been addressed in the field of astronomy with a focus on photometric data, see, e.g., the overview given by Ball and Brunner~\cite{BallB2009} and the references therein. To the best of our knowledge, the work reported upon in this paper is the first extensive machine-learning-based study on classifying quasars given spectroscopic data.


\section{Data}
\label{sec:data}
Our classification approaches work on a subset of the SDSS~(DR6)~\cite{sdss2010} database. This data has been obtained via a 2.5-meter telescope at the Apache Point Observatory (New Mexico) which is equipped with two special-purpose instruments: a 120-megapixel camera and a pair of spectrographs. Both types of resulting data, namely the photometric as well as the spectroscopic
data, are used in this work. 


\subsection{Labels}
To obtain ground truth data, we asked an expert to manually label (based on normalized plots) a random subset of $5,261$ spectra out of the $1,271,680$ spectra available in the SDSS.
The resulting catalog contains IDs and labels of $512$ objects of type ``quasar'' and $4,749$ objects of type ``other''.\footnote{The induced catalog along with the photometric and spectroscopic data can be obtained from the authors upon request.}
 
\subsection{Photometric Data}
The 120-megapixel camera collects simultaneously data through five different filters. By doing drift-scanning, those data sets are
combined to overlaying stripes that cover five wavelength ranges per observation. These wavelength ranges are the so-called \texttt{u},
\texttt{g}, \texttt{r}, \texttt{i},  and \texttt{z} \emph{bands}. In each of these bands, the SDSS photometric pipeline automatically extracts data for all celestial objects. Finally, in addition to the segmented raw data stripes, a batch of extracted photometric features is available for each object in each band.

One group of these photometric features, the so-called \emph{magnitudes}, are logarithmic measures of the brightness of an object. Among the most established approaches for data fitting are the \emph{PSF}, the \emph{Petrosian}, and the \emph{Model} approach~\cite{sdss2010}. The output of these approaches can be retrieved from the \texttt{PhotoObjAll} table of the \emph{Catalog Archive Server} (CAS)~\cite{sdss2010} and, since extracted from five bands, yields the following set of photometric features:

\begin{enumerate}
\item \raggedright{}PSF magnitudes: \ttt{psfMag\_u}, \ttt{psfMag\_g},
  \ttt{psfMag\_r}, \ttt{psfMag\_i}, \ttt{psfMag\_z}
\item \raggedright{}Petrosian magnitudes: \ttt{petroMag\_u}, \ttt{petroMag\_g},
  \ttt{petroMag\_r}, \ttt{petroMag\_i}, \ttt{petroMag\_z}
\item \raggedright{}Model magnitudes: \ttt{modelMag\_u}, \ttt{modelMag\_g},
  \ttt{modelMag\_r}, \ttt{modelMag\_i}, \ttt{modelMag\_z}
\end{enumerate}

\subsection{Spectroscopic Data}
\begin{figure}[h]
\begin{center}
  \resizebox{0.48\columnwidth}{!}{
    \includegraphics{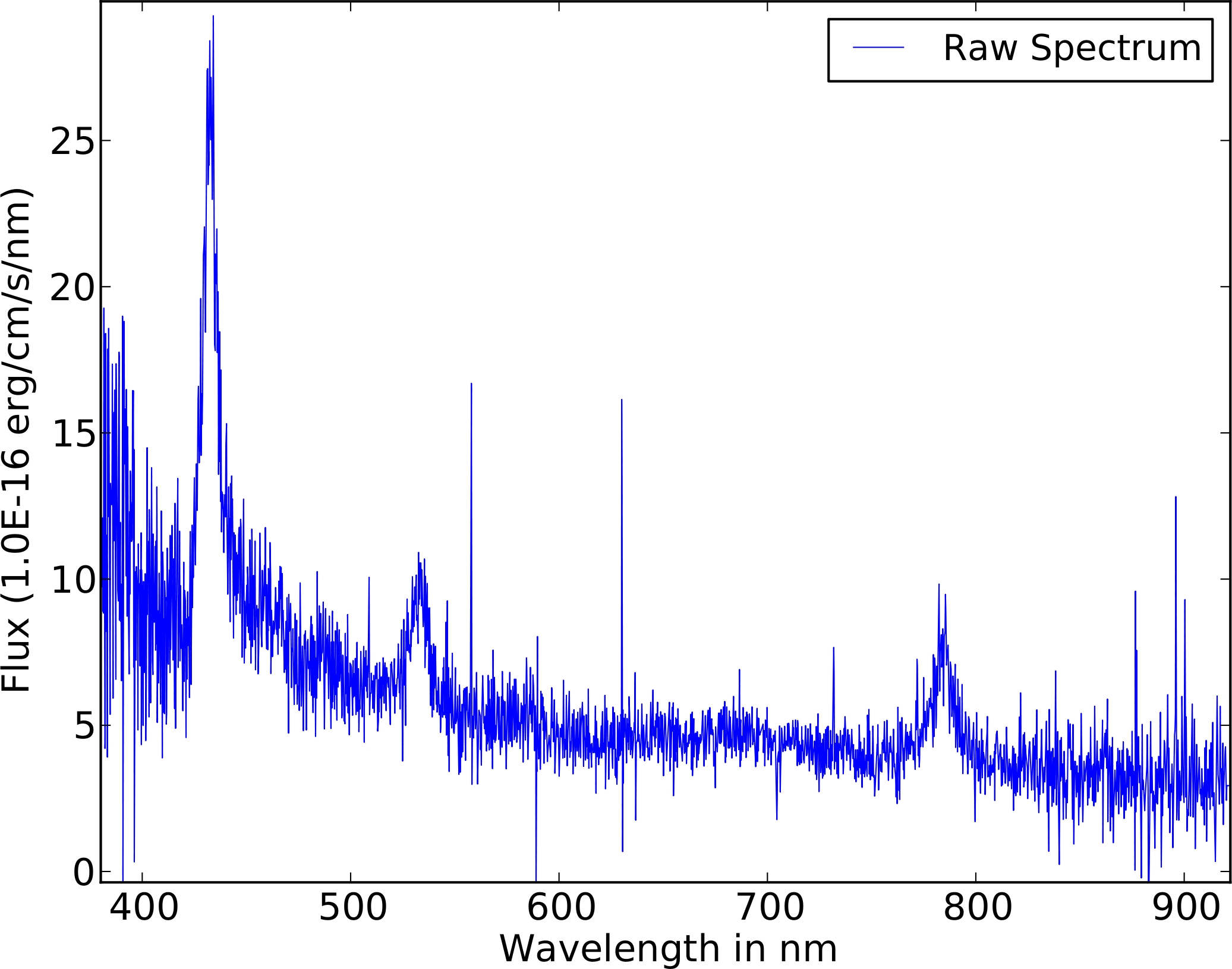}
  }\;\;
  \resizebox{0.48\columnwidth}{!}{
    \includegraphics{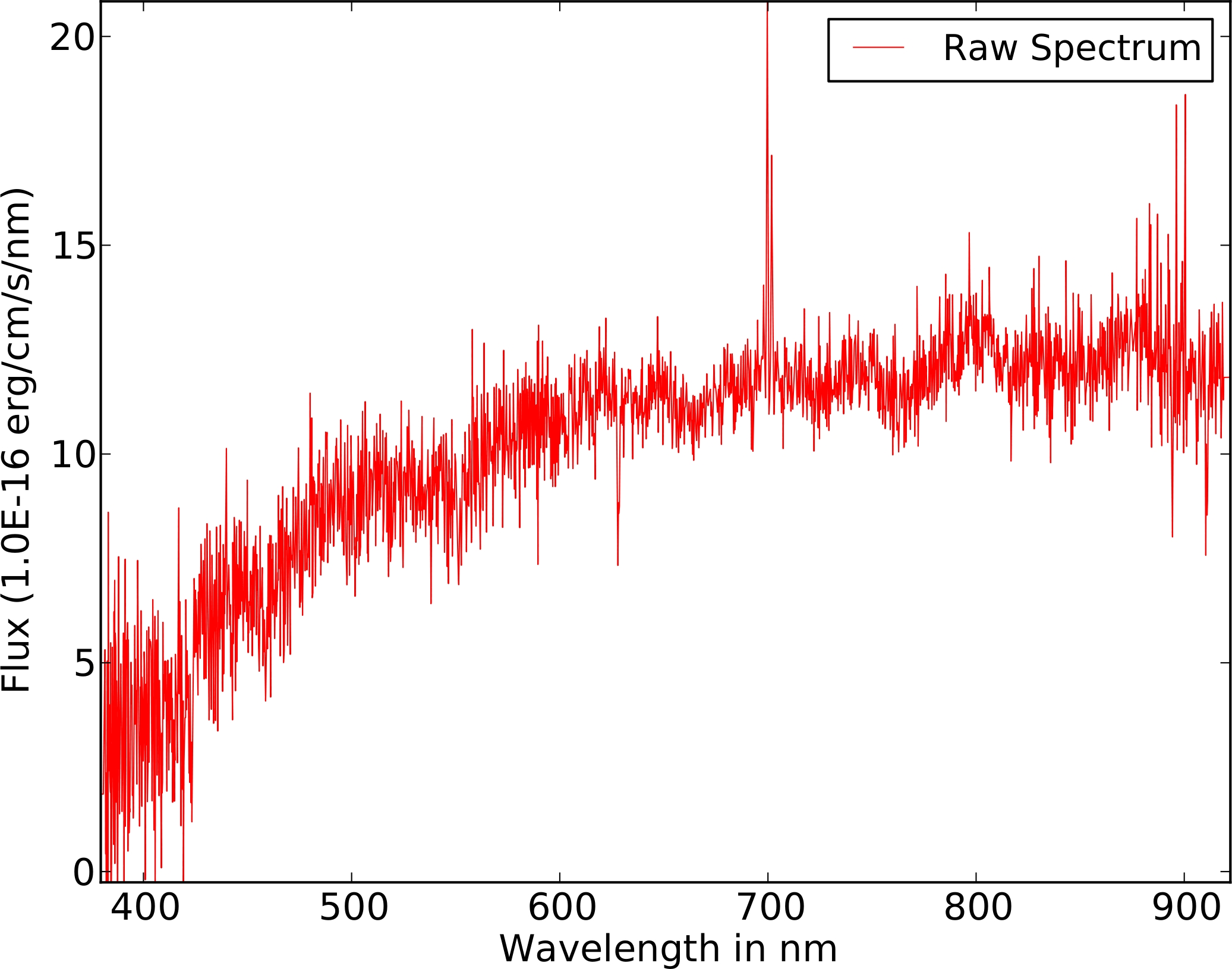}
  }
  \caption{Spectroscopic data for the manually labeled training set, which consists of objects of type ``quasars'' (blue) and of type ``other'' (red).}
  \label{fig:spec_data}
\end{center}
\end{figure}

The raw spectra of all manually labeled objects have been retrieved from the \emph{Data Archive Server} (DAS)~\cite{sdss2010}. Each
spectrum contains the flux values for roughly $3,850$ wavelengths. In order to obtain raw features that are independent of the specific
wavelengths, the spectra of all considered objects have been aligned and truncated. This results in a set of spectra having $d=3,825$ features, see Figure~\ref{fig:spec_data}.

\subsection{Quality of the Data Sample}
\label{sec:astronomical_comments}
\begin{figure}[h]
\begin{center}
  \resizebox{0.48\columnwidth}{!}{
    \includegraphics{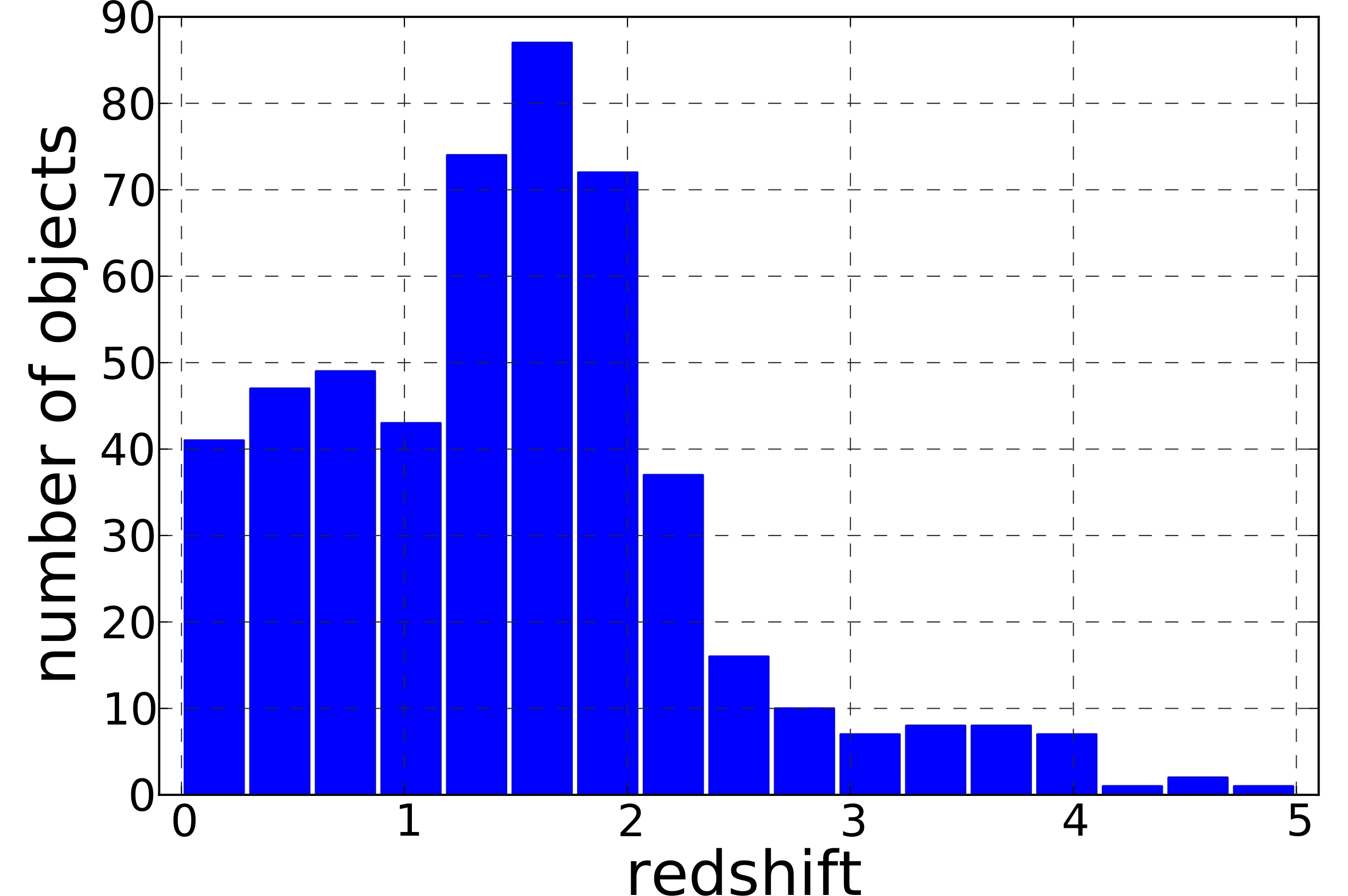}
  }\;
  \resizebox{0.48\columnwidth}{!}{
    \includegraphics{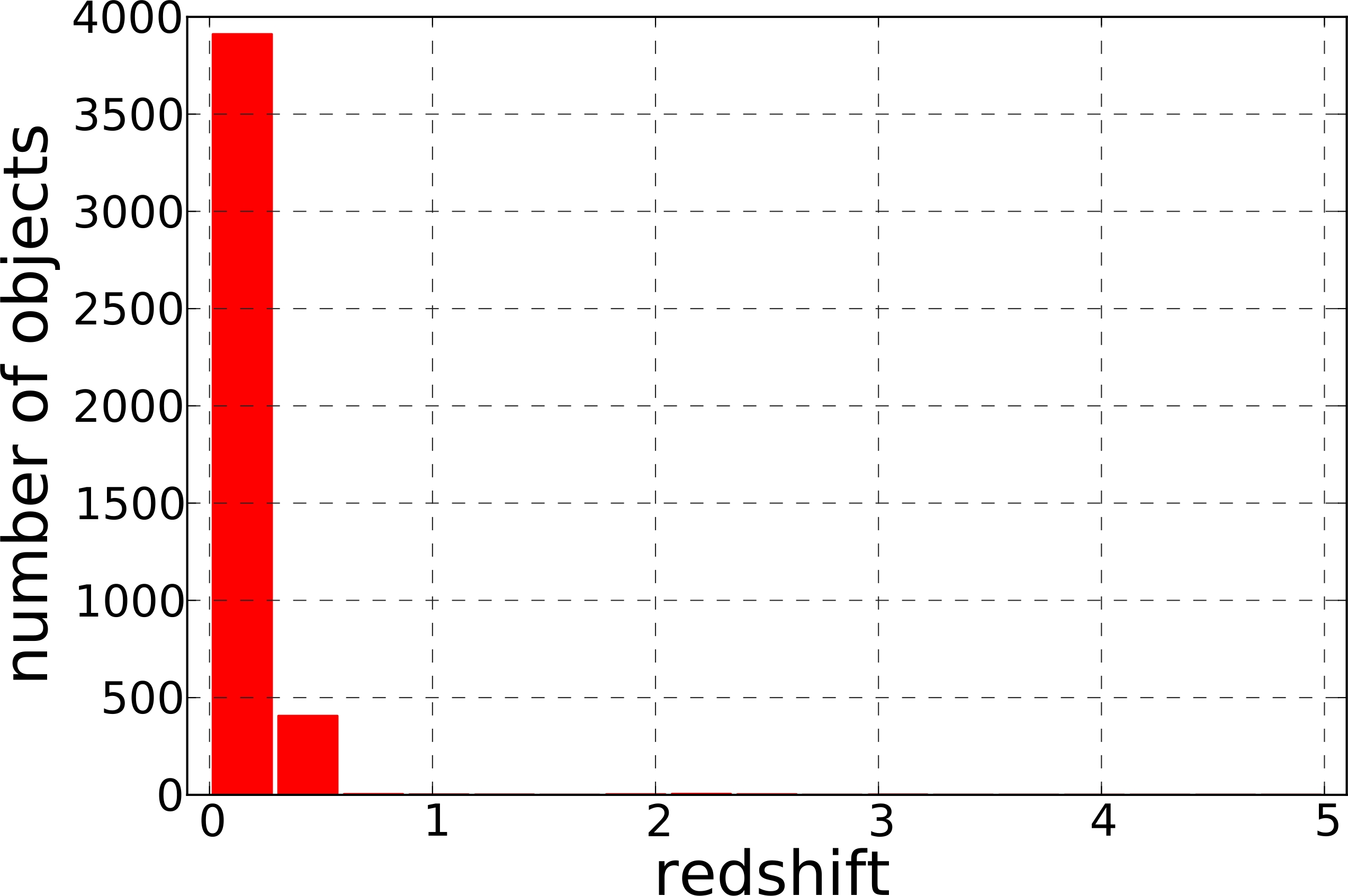}
  }\;
  \caption{Redshift distribution of quasars (left) and other objects (right).}
  \label{fig:redshifts}
\end{center}
\end{figure}

To estimate the quality of our data sample, we consider an important astronomical property, namely the \emph{redshift (value)} of an object~\cite{sdss2010}. In Figure~\ref{fig:redshifts}, the distribution (based on the $z$-values in the SDSS database) of all $512$ quasars present in our data set is given. It can be seen that these objects cover a redshift range of up to $z=5$ and that the majority has a low redshift \mbox{($z<2.3$)}. The remaining objects are dominated by $4,053$ galaxies exhibiting a low redshift.



\begin{figure}[h]
\begin{center}
  \resizebox{0.7\columnwidth}{!}{
    \includegraphics{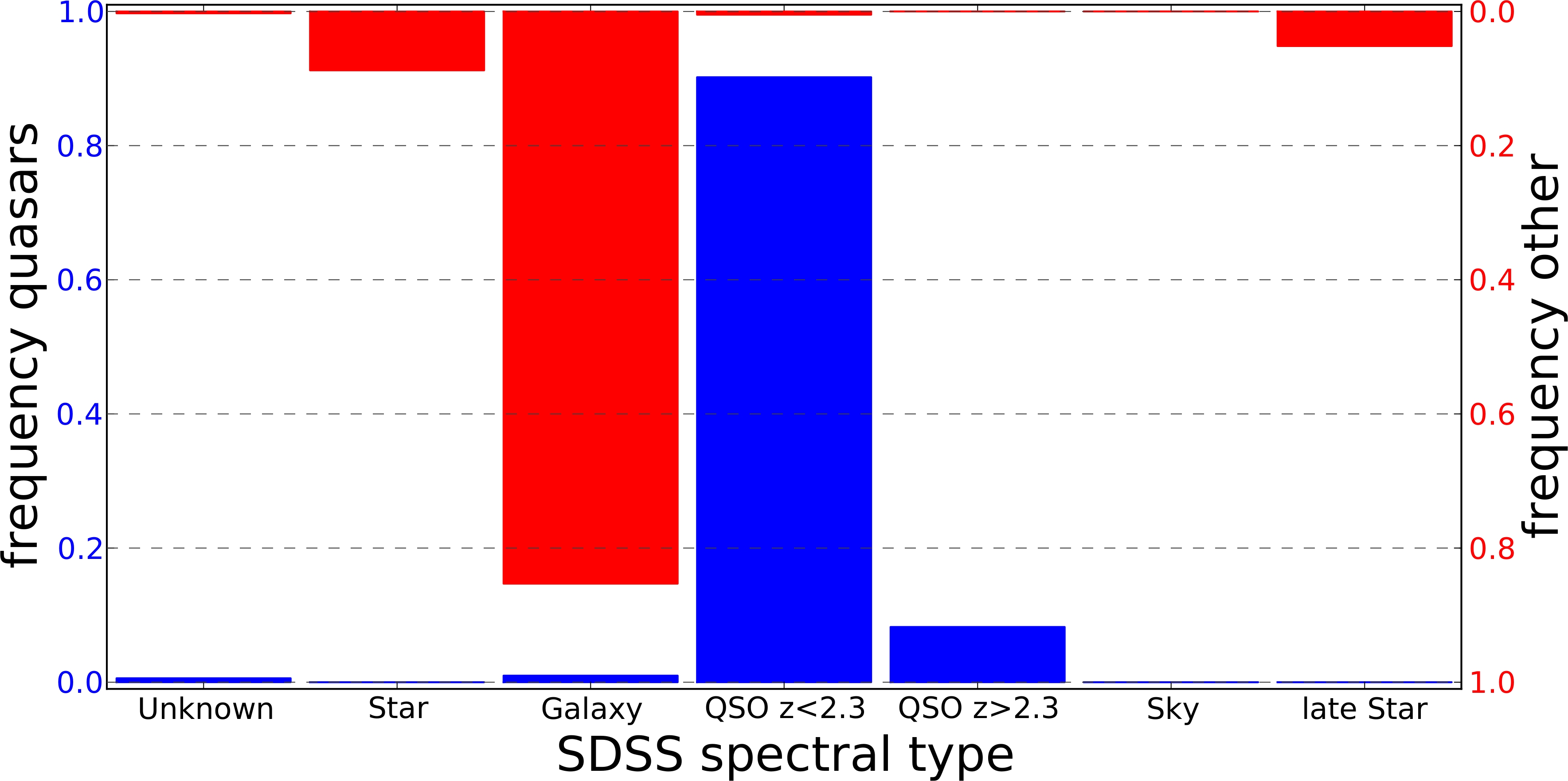}
  }
  \caption{Comparison of our expert's manual labeling with the corresponding labeling of the SDSS database.}
  \label{fig:types}
\end{center}
\end{figure}

In Figure~\ref{fig:types}, the labels available in the SDSS database are compared with our manually obtained labels. Here, the blue plot corresponds to all $512$ objects classified as objects of type ``quasar'' by our expert; the red plot corresponds to the remaining objects of type ''other''. The $x$-axis of both plots is based on the labels present in the SDSS database. It can be clearly seen that the labeling of our expert is in agreement with the one of the SDSS-pipeline.\footnote{It should be pointed out that the SDSS-pipeline is also based on human verification of the labels and, hence, can only be seen as a semi-automatic classification approach.} 




%

\section{Classification Approaches}
\label{sec:classification}
The labeled data induces a (training) set $\Tset = \{(\vec{x}_1,y_1),\ldots,(\vec{x}_n,y_\tsize)\} \subset \Reals^d \times \{-1,+1\}$ containing the features and labels for each object, where the positive class corresponds to all objects of type ''quasar''. We apply two well-known classification approaches, namely \emph{$k$-Nearest Neighbors (kNNs)} and \emph{Support Vector Machines (SVMs)}. For completeness, we depict the key ideas of both classification models and refer to the corresponding literature for details~\cite{HastieTF2009,SteinwartC2008,Vapnik1998}.
\subsection{$k$-Nearest Neighbors}
The $k$-Nearest Neighbor classifier uses the $k$ ``closest'' objects from the given set of already classified objects to assign a class to an unclassified object~\cite{HastieTF2009}. More precisely, the (binary) classification $\hat{Y}(\vec{x})$ for an object $\vec{x}$ is
\begin{equation} 
\label{eq:classification}
 \hat{Y}(\vec{x}) = 
\begin{cases}
 \phantom{-}1 & \text{if }  f(\vec{x}) > 0 \\
 -1 & \text{if } f(\vec{x}) \leq 0,
\end{cases}
\end{equation}
where
\begin{equation}
f(\vec{x}) = \sum_{\vec{x}_i \in N_k (\vec{x})} y_i,
\end{equation}
and where $N_k(\vec{x})$ denotes the $k$-nearest neighbors in the training set with respect to $\vec{x}$. To define ``closeness'', arbitrary metrics can be used; a popular choice is the Euclidean metric (which we use for the experimental evaluation as well).
\subsection{Support Vector Machines} 
The aim of a SVM consists in finding a hyperplane in a feature space which maximizes the ``margin'' between both classes such that only few training patterns lie within the margin~\cite{SteinwartC2008,Vapnik1998}. The latter task can be formulated as an optimization problem, where the first term corresponds to maximizing the margin and the second term to the loss caused by patterns lying within the margin:
\begin{center}

\begin{align}
\minimize_{\vec{w} \in \HilbertSpace_0,\;\vec{\xi} \in \Reals^\tsize,\;b\in\Reals} &\; \frac{1}{2} \norm{\vec{w}}^2 + C \sum_{i=1}^n \vec{\xi}_i\notag\\
\text{s.t.}                           &\; y_i(\ip{\vec{w}}{\Phi(\vec{x}_i)} + b) \geq 1 - \xi_i,\\  
	                           &\;\text{and } \xi_i \geq 0,\notag
\end{align} 

\end{center}
where $C > 0$ is a user-defined parameter. The function $\Phi:\Reals^d \rightarrow \HilbertSpace_0$ mapping the patterns into a feature space $\HilbertSpace_0$ is induced by a kernel function $\kernel:\Reals^d\times \Reals^d \rightarrow \Reals$ with $\kernel(\vec{x}_i,\vec{x}_j) = \ip{\Phi(\vec{x}_i)}{\Phi(\vec{x}_j)}$. A kernel function can be seen as a ''similarity measure'' for input patterns. The goal of the learning process is to find the optimal hyperplane $f(\vec{x}) = \ip{\vec{w}}{\Phi(\vec{x})} + b$. Unseen objects can subsequently be classified via Equation (\ref{eq:classification}). A common choice for the kernel function is the linear kernel 
\begin{equation}
 \kernel(\vec{x}_i,\vec{x}_j) = \ip{\vec{x}_i}{\vec{x}_j}
\end{equation}
or the RBF kernel
\begin{equation}
\kernel(\vec{x}_i,\vec{x}_j) = \exp \left(- \frac{{\|\vec{x}_i - \vec{x}_j\|}^2}{2 \sigma^2}\right), 
\end{equation}
where the parameter $\sigma$ is called the \emph{kernel width}.

\section{Spectroscopic Feature Extraction}
\label{sec:feature_extraction}
As discussed in Section~\ref{sec:background}, the physical properties of quasars result in characteristic, broad emission lines. Our approach consists in extracting meaningful features given ``continuum-substracted'' versions of the raw spectra. In the remainder of this section, we provide two ways for estimating the continuum (i.e. the rough shape) of a spectrum. Given such an estimate, we define several features motivated by the physical properties of quasars and non-quasars.




\subsection{Continuum Extraction with Splines}
\begin{figure*}[t]
\subfigure[Spline Model]{
\resizebox{0.85\columnwidth}{!}{
\includegraphics{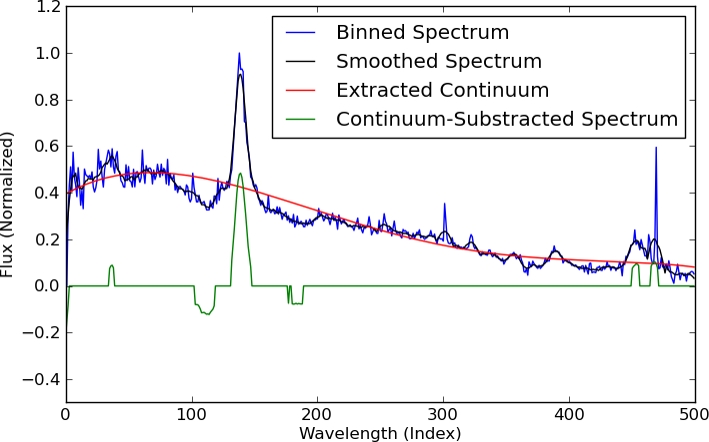}
}\;\;\;\;\;
}
\;\;\;\;\;\;\;\;\;\;\;\;\;\;\;\;\;
\subfigure[Support Vector Regression Model]{
\resizebox{0.85\columnwidth}{!}{
\includegraphics{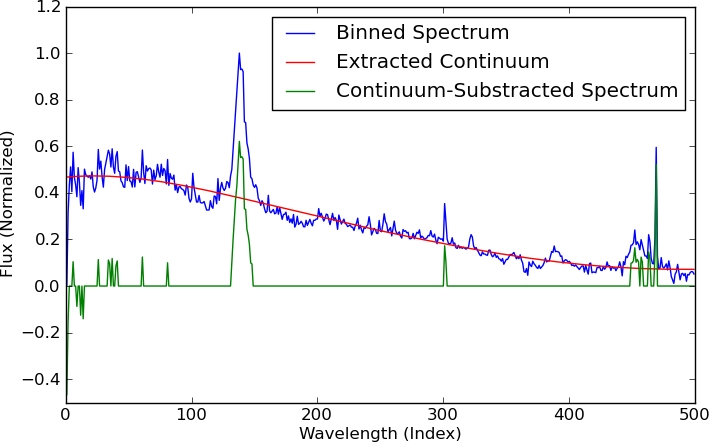}
}\;\;\;
}

\caption{Estimation of the continuum with splines and support vector regression.}
\label{fig:featureExtraction}
\end{figure*}
The first approach for extracting the continuum is based on splines~\cite{Hildebrand1987}. In a first step, we use a sliding-window technique to merge consecutive features (flux values) and to obtain a ``binned'' version of the spectroscopic data. This binned spectrum is subsequently normalized and smoothed using the Savitzky-Golay-Filter~\cite{Savitzky1964} (which is a standard smoothing filter), see the black curve shown in Figure~\ref{fig:featureExtraction}. Since the broad emission lines present in a quasar's spectroscopic data lead to an overestimation of the continuum, we apply an additional smoothing method to reduce the influence of the peak intensities. Given such a smoothed spectrum, we perform a spline interpolation to extract the estimated continuum (red curve). Finally, the difference of the smoothed spectrum and the estimated continuum is evaluated by rounding ``nonrelevant'' values within a $\delta$-range around the continuum, where $\delta$ is computed as the standard deviation of all calculated differences. The resulting ``continuum-subtracted'' spectrum is represented by the green curve in Figure~\ref{fig:featureExtraction}.

\subsection{Continuum Extraction with Support Vector Regression}
Our second approach for estimating the continuum is based on the concept of \emph{support vector regression (SVR)}~\cite{SteinwartC2008} which can be considered to be a special case of regularization problems of the form
\begin{equation}
\label{eq:tikhonov}
 \inf_{f \in \HilbertSpace} \; \frac{1}{\tsize} \sum_{i=1}^n \loss \big( y_i,f(\vec{x}_i) \big) + \lambda \norm{f}_\HilbertSpace^2,
\end{equation}
where $\lambda > 0$ is a fixed real number, $\loss:\Reals \times \Reals \rightarrow [0,\infty)$ is a \emph{loss function} and $\norm{f}_\HilbertSpace^2$ is the squared norm in a \emph{reproducing kernel Hilbert space} (RKHS) $\HilbertSpace \subseteq \Reals^X = \{f: X \rightarrow \Reals\}$ induced by the associated kernel function (note that, in this case, the kernel function is defined on the indices associated with the flux vales). By plugging in the so-called $\varepsilon$-\emph{intensive loss} with $\varepsilon>0$ one obtains
\begin{equation}
\label{eq:svr}
 \inf_{f \in \HilbertSpace} \; \frac{1}{\tsize} \sum_{i=1}^n  \max ( |f(\vec{x}_i)-y_i|-\varepsilon,0 \big) + \lambda \norm{f}_\HilbertSpace^2,
\end{equation}
which depicts the optimization problem of SVR. Here, the first term corresponds to the ``difference'' between the (continuum) model and the data and the second term corresponds to the ``complexity'' of the (continuum) model. Ideally, one would like to have a model which fits the data well and which is not too ``complex''. Additionally (and in contrast to standard regression problems), we would like to ``ignore'' the broad emission lines present in the quasars' spectra, see Figure~\ref{fig:featureExtraction}. Two properties of the SVR approach are advantageous in this context: First, the $\varepsilon$-intensive loss function penalizes the differences of the model and the flux values only linearly (in contrast to, e.g., the square-loss~\cite{SteinwartC2008}). Second, by adapting the involved parameters, one can adjust the behavior of the model such that the influence of large peaks is reduced.\footnote{For instance, the kernel width $\sigma$ of the (used) RBF kernel can be adjusted such that large peaks have less influence on the continuum model.} 

Given the resulting continuum models, we compute the continuum substracted spectra by considering the difference between each continuum model and the corresponding input spectrum while ignoring values lying within a small $\delta$-range around the continuum model (like above).

\subsection{Extracted Features}
Given the estimated continuum and thus the continuum-substracted spectra, we try to extract meaningful features representing the main characteristics of typical quasars, see Table~\ref{table:features}: Since the (true) continuum of a quasar's spectrum is normally horizontal or decreasing, we use the first and the last value of the computed continuum as features to estimate the rough slope of the spectrum (F1 and F2). As described in Section~\ref{sec:background}, broad emission lines provide an excellent characteristic for experts to classify quasars. The remaining eight features aim at capturing these ``peak properties''. Here, features F3 and F4 denote the absolute values of the sum of all ``positive'' and ``negative'' peaks, respectively. Further, features F5, F6, F9, and F10 indicate the width and the face of the strongest peaks. Finally, features F7 and F8 capture the minimum and maximum peak intensities.


\begin{table}[t]
\begin{center}
\caption{Extracted Features}
\begin{tabular}{|c|p{0.333\textwidth}|}
\hline
Feature & Description  \\\hline\hline
F1 &  \ttt{First value of the extracted continuum}\\\hline
F2 &  \ttt{Last value of the extracted continuum}\\\hline
F3 &  \ttt{Integral of all positive peaks}\\\hline
F4 &  \ttt{Integral of all negative peaks}\\\hline
F5 &  \ttt{Width of the broadest positive peak}\\\hline
F6 &  \ttt{Width of the broadest negative peak}\\\hline
F7 &  \ttt{Major peak intensity}\\\hline
F8 &  \ttt{Minor peak intensity}\\\hline
F9 &  \ttt{Major face of positive peak} \\\hline
F10 & \ttt{Major face of negative peak}\\\hline
\end{tabular} 
\label{table:features}
\end{center} 
\end{table}

\section{Experiments}
\label{sec:experiments}
We consider several data sets which are based on the photometric and spectroscopic data in order to evaluate the generalization performances of our approaches. In the remainder of this section, we provide a description of these data sets as well as a description of the experimental setup and the final outcome of the experimental evaluation.
\subsection{Data Sets} Given the data, we generate a variety of data sets, see Table~\ref{table:data_sets}. Each data set contains all $N=5,261$ objects and, thus, all $p=512$ objects of type ``quasar'' and all $n=4,749$ objects of type ``other''. The first four data sets D1--D4 are based on the different magnitudes which are retrieved from the photometric data, see Section~\ref{sec:data}. The data sets D5--D7 are binned versions of the raw spectra, i.e., the original input dimension ($3,825$) of each spectrum is reduced to $d=5,d=100$ and $d=500$ respectively by ``averaging'' consecutive flux values such that the desired output dimension is obtained. The data sets $D8$ and $D9$ are obtained via the spectroscopic feature extraction approaches depicted in Section~\ref{sec:feature_extraction}, where we use binned versions of the spectra with $d=500$ dimensions as starting point. The last data set $D10$ is based on the spectroscopic feature extraction (SVR) in combination with a set of photometric features.\footnote{This data set aims at the situation, where both parts of the data (photometric and spectroscopic) are given for all objects (which is the case for, e.g., the LAMOST project~\cite{lamost2010}).}



\begin{table}[t]
\begin{center}
\caption{Considered Data Sets}
\newcolumntype{P}[1]{>{#1\arraybackslash}p{0.375\textwidth}}
\begin{tabular}{|c|P{\raggedright}|}
\hline
Data Set & Features  \\\hline\hline
D1 & \ttt{psfMag\_u} - \ttt{psfMag\_g}, \ttt{psfMag\_g} - \ttt{psfMag\_r}, \ttt{psfMag\_r} - \ttt{psfMag\_i}, \ttt{psfMag\_i} - \ttt{psfMag\_z} \\\hline
D2 & \ttt{psfMag\_u}, \ttt{psfMag\_g}, \ttt{psfMag\_r}, \ttt{psfMag\_i}, \ttt{psfMag\_z} \\\hline
D3 & \ttt{psfMag\_u}, \ttt{psfMag\_g}, \ttt{psfMag\_r}, \ttt{psfMag\_i}, \ttt{psfMag\_z}, \ttt{modelMag\_u}, \ttt{modelMag\_g}, \ttt{modelMag\_r}, \ttt{modelMag\_i}, \ttt{modelMag\_z} \\\hline
D4 & \ttt{psfMag\_u}, \ttt{psfMag\_g}, \ttt{psfMag\_r}, \ttt{psfMag\_i}, \ttt{psfMag\_z}, \ttt{petroMag\_u}, \ttt{petroMag\_g}, \ttt{petroMag\_r}, \ttt{petroMag\_i}, \ttt{petroMag\_z} \\\hline
D5 & \ttt{BinnedSpec5} \\\hline
D6 & \ttt{BinnedSpec100} \\\hline
D7 & \ttt{BinnedSpec500} \\\hline
D8 & \ttt{ExtractedFeatures (SPLINE)} \\\hline
D9 & \ttt{ExtractedFeatures (SVR)} \\\hline
D10 & \ttt{ExtractedFeatures (SVR)}, \ttt{psfMag\_u}, \ttt{psfMag\_g}, \ttt{psfMag\_r}, \ttt{psfMag\_i}, \ttt{psfMag\_z}, \ttt{modelMag\_u}, \ttt{modelMag\_g}, \ttt{modelMag\_r}, \ttt{modelMag\_i}, \ttt{modelMag\_z}  \\\hline
\end{tabular} 
\label{table:data_sets}
\end{center} 
\end{table}

\subsection{Experimental Setup} 
Except for the SVM implementation, all preprocessing steps and
algorithms are implemented in \texttt{Python}. For the SVM model, we resort to the \texttt{LIBSVM} implementation provided by Chang and Lin~\cite{libsvm2010}.
\paragraph{Parameters} 
To train and evaluate the classification approaches, half of each data set is used as training and the other half as test set. For the SVM model, a RBF kernel with kernel width $\sigma$ is used. The corresponding model parameters $k, C,$ and $\sigma$ for both classification approaches are tuned via 10-fold cross-validation~\cite{HastieTF2009} on the training set given a grid of parameters ($k \in \{1,\ldots,10\}$ for kNN and $(C,\sigma) \in \{2^{-10},2^{-9}\ldots,2^{+10}\} \times \{2^{-10},2^{-9},\ldots,2^{+10}\}$ for the SVM). For the spectroscopic feature extraction, we set the involved parameters to fixed values for all spectra: For the spline-based approach, the binned spectrum is smoothed using the Savitzky-Golay-filter of degree $3$ and window size $15$. Afterwards we use the interpolation package of the \texttt{Scipy} library \cite{Scipy2001} for estimating the smoothed spline of order $4$ and smoothing factor $100$. For the SVR-based approach, we again make use of the \texttt{LIBSVM} implementation using a RBF kernel and input parameters $\texttt{C}=100$, $\texttt{gamma}=0.00001$, and $\texttt{epsilon}=0.1$.
\paragraph{Measuring the Classification Performance}
The performances of our models are evaluated on the test set. Since our data sets are imbalanced, we resort to the \emph{Matthews Correlation Coefficient} (MCC)~\cite{Matthews1975} as quality measure:
\begin{equation*}
\frac{\ensuremath{\mathit{TP}} \cdot \ensuremath{\mathit{TN}} - \ensuremath{\mathit{FP}} \cdot \ensuremath{\mathit{FN}}}{\sqrt{(\ensuremath{\mathit{TP}}+\ensuremath{\mathit{FN}})(\ensuremath{\mathit{TP}}+\ensuremath{\mathit{FP}})(\ensuremath{\mathit{TN}}+\ensuremath{\mathit{FP}})(\ensuremath{\mathit{TN}}+\ensuremath{\mathit{FN}})}}
\end{equation*}
Here, $\ensuremath{\mathit{TP}},\ensuremath{\mathit{FP}},\ensuremath{\mathit{TN}}$ and $\ensuremath{\mathit{FN}}$ denote the number of true positives, false positives, true negatives, and false negatives, respectively~\cite{Matthews1975,Alpaydin2010}. We also provide the \emph{true positive rate (\ensuremath{\mathit{TP}}-rate)} given by $\ensuremath{\mathit{TP}}/p$ as well as the \emph{false positive rate (\ensuremath{\mathit{FP}}-rate)} given by $\ensuremath{\mathit{FP}}/n$, which can be seen as ``hit rate'' and ``false alarm rate'', respectively~\cite{Alpaydin2010}. Finally, we consider the \emph{error} on the test set given by $(\ensuremath{\mathit{FP}}+\ensuremath{\mathit{FN}})/N$ which simply measures the overall number of misclassifications. 
\subsection{Results}
In the remainder of this section, we discuss the outcome of the conducted experiments.
\paragraph{Classification Performance}
\begin{table}[t]
\begin{center}
\caption{Classification Performances}
\begin{tabular}{|c||c|c|c|c|}
\hline
Data & \multicolumn{4}{c|}{kNN} \\\cline{2-5}
Set & MCC & Error & \ensuremath{\mathit{TP-rate}} & \ensuremath{\mathit{FP-rate}}\\\hline\hline
D1 & 0.873 & 2.32\% & 86.1\% & 1.0\% \\\hline
D2 & 0.826 & 3.15\% & 81.7\% & 1.4\% \\\hline
D3 & 0.902 & 1.82\% & 91.2\% & 1.0\% \\\hline
D4 & 0.893 & 2.01\% & 91.6\% & 1.3\% \\\hline
D5 & 0.748 & 4.41\% & 70.6\% & 1.5\% \\\hline
D6 & 0.824 & 3.15\% & 80.2\% & 1.2\% \\\hline
D7 & 0.808 & 3.42\% & 77.7\% & 1.2\% \\\hline
D8 & 0.963 & 0.69\% & 95.6\% & 0.3\% \\\hline
D9 & 0.959 & 0.76\% & 96.7\% & 0.5\% \\\hline
D10 & 0.971 & 0.53\% & 96.3\% & {\bf 0.2\%} \\\hline
\end{tabular}

\vspace{5ex}

\begin{tabular}{|c||c|c|c|c|}
\hline
Data & \multicolumn{4}{c|}{SVM (RBF)}\\\cline{2-5}
Set & MCC & Error & \ensuremath{\mathit{TP-rate}} & \ensuremath{\mathit{FP-rate}}\\\hline\hline
D1 & 0.859 & 2.54\% & 84.2\% & 1.0\%\\\hline
D2 & 0.851 & 2.74\% & 85.0\% & 1.3\%\\\hline
D3 & 0.922 & 1.44\% & 91.9\% & 0.7\%\\\hline
D4 & 0.908 & 1.71\% & 91.5\% & 0.9\%\\\hline
D5 & 0.757 & 4.33\% & 73.6\% & 1.8\%\\\hline
D6 & 0.788 & 3.69\% & 71.7\% & 0.8\%\\\hline
D7 & 0.610 & 6.23\% & 44.0\% & 0.5\%\\\hline
D8 & 0.965 & 0.65\% & 96.0\% & 0.3\%\\\hline
D9 & {\bf 0.977} & {\bf 0.42\%} & {\bf 97.4\%} & {\bf 0.2\%}\\\hline
D10 & 0.971 & 0.53\% & 96.7\% & {\bf 0.2\%}\\\hline
\end{tabular}
\label{tab:classification_performances}
\end{center}
\end{table}
The classification performances of both classification approaches on the various data sets are shown in Table~\ref{tab:classification_performances}, where the best results are highlighted (bold face type). Both the kNN and the SVM model perform well on the first four data sets, which are composed of the magnitudes retrieved from the photometric data.\footnote{We would like to point out that a generalization of these results to the complete photometric catalog is dubious due to the fact that the sample is biased (a target selection criteria was used for collecting the spectroscopic data~\cite{sdss2010}).} While the performance is worse on the data sets containing the binned versions of the spectra (D5--D7), it improves significantly when the extracted features (D8 and D9) are used as input data. Using additional photometric features (D10) does not lead to an imrovement.


\paragraph{How Much Spectroscopic Data is Needed}
The spectroscopic feature extraction approaches used for the data sets D8--D10 are based on binned versions of the raw spectra containing $d=500$ values. A natural question is ``how detailed'' these binned versions have to be such that the spectroscopic feature extraction yields good results. One of the benefits of being able to use ``low detailed'' versions of the raw spectra is the performance gain with respect to the computation time (which will be of tremendous importance for an efficient classification of larger portions of the SDSS or even the full catalog).

\begin{figure}[t]
\begin{center}
\resizebox{0.8\columnwidth}{!}{
\includegraphics{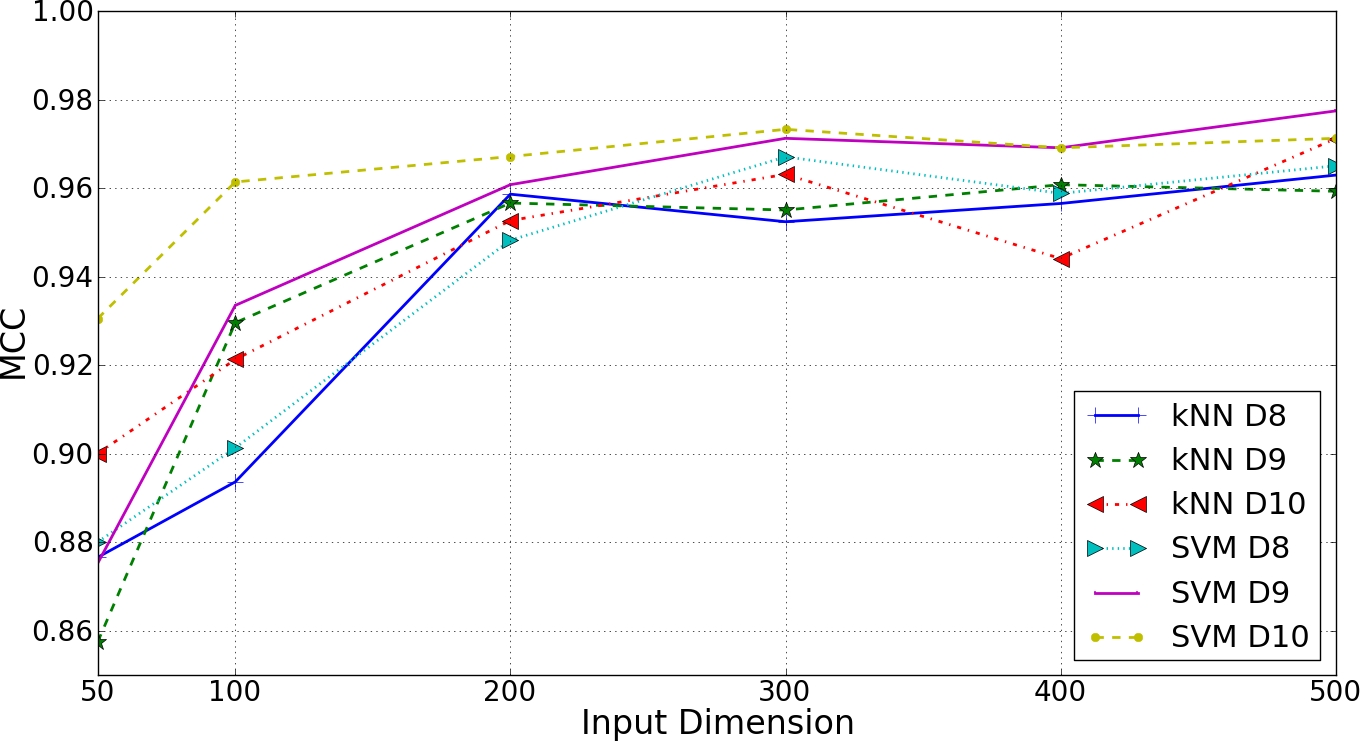}
}\;
\end{center}
\caption{Influence of the input dimension on the MCC.}
\label{fig:howmuchdata_MCC}
\end{figure}

In Figure~\ref{fig:howmuchdata_MCC}, the influence of the input dimension on the classification performance (MCC) is shown. The plot indicates a quite similar classification performance of both approaches on all considered data sets. Further, it seems that data sets based on input dimensions of roughly $d=200$ already yield good results, which cannot be improved significantly by using a higher input precision. We expect this observation to be helpful when considering a time/classification quality trade-off for processing massive databases, e.g. in the context of the LAMOST project~\cite{lamost2010}.


\section{Conclusions and Outlook}
We have considered the task of using classification algorithms to discriminate quasars from other objects given photometric and spectroscopic data. In recent years, problems of this kind have gained more and more attention in the field of astronomy due to the fact that the massive amounts of data arising nowadays cannot be handled without automatic (machine learning) approaches. While the problem of detecting quasars (especially using only photometric data) has already been considered in the field of astronomy, we are not aware of corresponding literature (with respect to spectroscopic data) in the field of machine learning, and the orthogonality of the existing and proposed approaches seems to allow for a mutually beneficial collaboration between these fields. The shear volume of astronomical data that is already available or will be available in the near future emphasizes the need for high-performance machine learning algorithms.


\bibliographystyle{plain}
\bibliography{paper_astro-ph}

\end{document}